\documentclass[11pt,a4paper]{article}
\usepackage{amsmath}
\usepackage{amsfonts}
\usepackage{amssymb}
\usepackage[left=1.5cm, right=1.5cm, top=1.50cm, bottom=1.50cm]{geometry}
\usepackage{natbib}
\usepackage{booktabs}
\usepackage{setspace}
\usepackage{subfigure}
\usepackage{graphicx}
\usepackage{fancyhdr}
\usepackage{rotating}
\usepackage{array}
\usepackage{color}
\usepackage{multirow}
\usepackage{float}
\usepackage{url}
\usepackage{subfigure}
\usepackage{hyperref}
\usepackage{abstract}
\usepackage{multicol}
\hypersetup{
unicode=false,          % non-Latin characters in Acrobat’s bookmarks
pdftoolbar=true,        % show Acrobat’s toolbar?
pdfmenubar=true,        % show Acrobat’s menu?
pdffitwindow=false,     % window fit to page when opened
pdftitle={Dayglow on Mars},    % title
pdfauthor={Sonal Kumar Jain},   % author
pdfsubject={Cameron band and UV doublet band},   % subject of the document
colorlinks=true,        % false: boxed links; true: colored links
linkcolor=blue,         % color of internal links
citecolor=blue,        % color of links to bibliography
filecolor=magenta,      % color of file links
}

\newcommand{\car}{{$\mathrm{CO_2} $}}
\newcommand{\carp}{{$\mathrm{CO_2^+} $}}

\newcommand{\nt}{{$\mathrm{N_2}$}}
\newcommand{\dgr}{{$^\circ$}}

\title{\bf Calculations of N$_2$ triplet states vibrational populations and band emissions in  Venusian dayglow}
\date{}
\author{Anil Bhardwaj\thanks{anil\_bhardwaj@vssc.gov.in; bhardwaj\_spl@yahoo.com} and 
Sonal Kumar Jain\thanks{sonaljain.spl@gmail.com}  \\ 
Space Physics Laboratory,\\ 
Vikram Sarabhai Space Centre,\\ 
Trivandrum, India - 695022}

\begin{document}

\maketitle

\begin{abstract}
A model for N$_2$ triplet states band emissions in the Venusian dayglow 
has been developed for low and high solar activity conditions.
Steady state photoelectron fluxes and  volume excitation rates for \nt\ 
triplet states have been calculated using the Analytical Yield Spectra (AYS) 
technique. Model calculated photoelectron flux is in good agreement with 
Pioneer Venus Orbiter-observed electron flux. Since inter-state 
cascading is important for the triplet states of \nt,  populations
of different levels of \nt\ triplet states are calculated under statistical
equilibrium considering direct electron impact excitation, and cascading and 
quenching effects. Densities of all vibrational levels of
each triplet state are calculated in the model. Height-integrated overhead
intensities of N$_2$ triplet band emissions are calculated, the values for
Vegard-Kaplan ($ A^3\Sigma_u^+ - X^1\Sigma^+_g $),
First Positive ($ B^3\Pi_g - A^3\Sigma^+_u $), Second Positive ($ C^3\Pi_u -
B^3\Pi_g $), and Wu-Benesch ($W^3\Delta_u - B^3\Pi_g$) bands of $N_2$, are 1.9 (3.2),
3 (6), 0.4 (0.8), and 0.5 (1.1) kR, respectively, for solar minimum (maximum)
conditions. The intensities of the three strong Vegard-Kaplan bands (0, 5), (0, 6),
and (0, 7) are 94 (160), 120 (204), and 114 (194) R, respectively, for solar minimum
(maximum) conditions. Limb profiles are calculated for VK (0, 4), (0, 5), (0, 6) and
(0, 7) bands.  The calculated intensities on Venus are about a factor 10 higher than
those on Mars. The present study provides a motivation for a search of N$_2$ triplet
band emissions in the dayglow of Venus.
\end{abstract}

\begin{multicols}{2}
\section{Introduction}

Ultraviolet (UV) emissions of Venus have been studied for decades by
rocket-borne spectrometers, Mariner 10, Venera 11 and 12, and Galileo 
spacecraft flybys, Pioneer Venus Orbiter, and Hopkins Ultraviolet Telescope
aboard Space Shuttle \citep[e.g.,][]{Fox91,Paxton92,Feldman00,Gerard08}.
More recently, Venusian UV emissions observed during the Cassini flyby have been reported
\citep{Gerard11,Hubert10}. The major UV emissions observed are from H, He, C, N, O
and CO.  However, until now the \nt\ emissions have not been 
observed on Venus. Recently, SPICAM (Spectroscopy for Investigation of
Characteristics of the Atmosphere of Mars) onboard Mars Express
has observed, for the first time, emissions from \nt\ triplet excited states. 
The main emissions observed are the (0, 6) and (0, 5) bands of the
Vegard-Kaplan (VK) transitions of \nt\ \citep{Leblanc06,Leblanc07}.
On Venus, the VK bands would be more intense compared to those on Mars
because of its proximity to the Sun and higher \nt\ abundances. 
SPICAV (Spectroscopy for Investigation of Characteristics of the Atmosphere of Venus)
on Venus Express orbiter mission, which is similar to SPICAM, can able to
observe these emissions in the Venusian dayglow.

We have recently developed  a model for
the \nt\ triplet state dayglow emissions on Mars using the Analytical 
Yield Spectra approach \citep{Jain11}. While calculating the emissions of 
triplet transitions of \nt, cascading from the higher lying states and
quenching by atmospheric constituents are considered and 
the population of any given vibrational level of a
state is calculated under statistical equilibrium.
In the present paper, this model is used to calculate the \nt\ dayglow 
emission on Venus for solar maximum and minimum conditions. 
Previous theoretical study on \nt\ emissions in the Venusian dayglow have been
carried out by \cite{Fox81} and \cite{Gronoff08}.  We compare our results 
with their model intensities.

\section{Model Input Parameters}
The model atmosphere considering five gases (\car, CO, \nt, O, and O$_2$)
is taken from the VTS3 model of \cite{Hedin83} for the low 
(F10.7 = 80) and high (F10.7 = 200) solar activity  conditions. The EUVAC model 
of \cite{Richards94} has been used to calculate the 37-bin solar EUV flux, 
which is based on the F10.7 solar index. The
EUVAC solar spectrum is scaled for the heliocentric distance of Venus (0.72 AU).
All calculations are made at solar zenith angle (SZA) of 60\dgr\ unless 
otherwise noted. Photoionization and photoabsorption 
cross sections for gases considered in the present study have been taken
from \cite{Richards94} and  \cite{Schunk00}. The branching ratios for excited
states of \carp, CO$^+$, N$_2^+$, O$^+$, and O$_2^+$  have been taken from
\cite{Avakyan98}. Franck-Condon factors and transition probabilities required
for calculating the intensity of a specific band $ \nu'-\nu'' $ of \nt\ are taken
from \cite{Gilmore92}.

The  electron impact \nt\ triplet states excitation cross sections are taken
from \cite{Itikawa06}, which have been fitted analytically \citep[cf.][]{Jackman77,
Bhardwaj09} and fitting parameters are given in \cite{Jain11}. Electron impact cross
sections for \car\ are  from \cite{Bhardwaj09} and for other gases from \cite{Jackman77}.

\section{Photoelectron Flux}
To calculate the photoelectron flux we have adopted the Analytical Yield 
Spectra (AYS) technique \citep[cf.][]{Bhardwaj90b,Singhal91,
Bhardwaj99a,Bhardwaj03,Bhardwaj99d}. The AYS is an analytical representation of 
numerical yield spectra obtained using the Monte Carlo model 
\citep[cf.][]{Singhal80,Singhal91,Bhardwaj99d,Bhardwaj09}.
Recently, the AYS model for electron degradation in \car\ 
has been developed by \cite{Bhardwaj09}. Further details of the AYS technique are
given in \cite{Bhardwaj09} and references therein.  Using AYS the 
photoelectron flux has been calculated as 
\begin{equation}\label{eq:a}
\phi(Z,E)=\int_{W_{kl}}^{100} \frac{Q(Z,E) U(E,E_0)}
{{\displaystyle\sum_{l}} n_l(Z)\sigma_{lT}(E)} \ dE_0
\end{equation}
where $\sigma_{lT}(E)$ is the total inelastic cross section for the
$l$th gas with density  $n_l$, and $U(E,E_0)$ is the two-dimensional AYS,
which embodies the non-spatial information of degradation process. It 
represents the equilibrium number of electrons per unit energy at an
energy $E$ resulting from the local energy degradation of an incident
electron of energy $E_0$. For the \car\ gas it is given as \citep{Bhardwaj09}
\begin{equation}\label{eq:b}
      U(E,E_0)=A_1E_k^s+A_2(E_k^{1-t}/\epsilon^{3/2 +r})+ 
      \frac{E_0B_0e^{x}/B_1}{(1+e^{x})^2}
\end{equation}
Here $E_k=E_0/1000$, $\epsilon=E/I$ ($I$ is the lowest ionization
threshold), and $x=(E-B_2)/B_1$. $A_1=0.027,\ A_2=1.20,\ t=0,\ 
r=0$, $s=-0.0536$, $B_0=10.095$, $B_1=5.5$, 
and $B_2=0.9$ are the best fit parameters.

For other gases, viz., O$_2$, \nt, O, and CO, we have used the AYS
given by \cite{Singhal80}
\begin{equation}\label{eq:d}
      U(E,E_0)=C_0+C_1(E_k+K)/[(E-M)^2+L^2].  %equation 17
\end{equation}
Here $C_0$, $C_1$, $K$, $M$, and $L$ are the
fitted parameters which are independent of the energy, and 
whose values are given by \cite{Singhal80}. The term $Q(Z,E)$ in 
equation~(\ref{eq:a}) is the primary photoelectron production rate 
\citep[cf.][]{Michael97,Bhardwaj03,Jain11}.

\begin{figure}
\centering
\includegraphics[width=30pc]{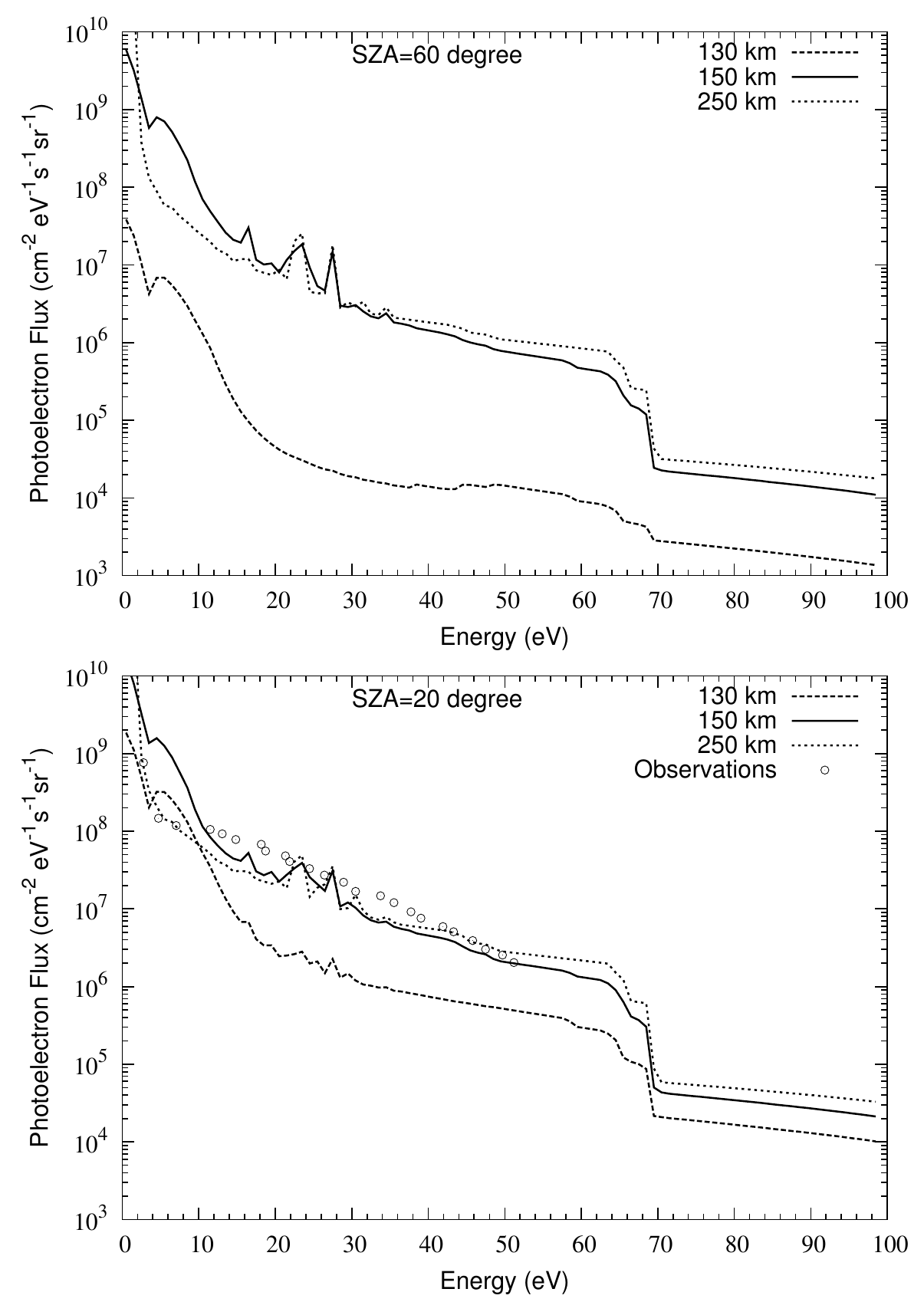}
\caption{Model calculated photoelectron flux for low (upper panel) and high (bottom panel)
solar activity conditions at 130, 150, and 230 km. Symbols in bottom panel represent
the Pioneer Venus Orbiter-observed values averaged over 206-296 km and 8\dgr-35\dgr\ SZA, taken from \cite{Spenner97}.}
\label{fig:pef}
\end{figure}

Model calculated photoelectron fluxes at 130, 150, and 250 km altitudes for
solar minimum condition at SZA=60\dgr, are shown in the upper panel of Figure~\ref{fig:pef}.
A sudden drop in flux at $ \sim $3 eV is due to the large vibrational cross sections 
at 3.8 eV for electron impact on \car\ \citep[cf.][]{Bhardwaj09}.
The sharp peak at 27 eV is due to ionization of \car\ in the ground state by 
the He II solar Lyman $ \alpha $ line at 303.78 \AA. The broad peak at 21-23 eV is 
due to ionization of \car\ in the A$^2\Pi_u$ and B$^2\Sigma_u^+$ states of  \carp\ 
by the 303.78 \AA\ solar photons.
Peak structures around 20-30 eV are clearly seen in the photoelectron
flux at 150 km, whereas they are smoothed out at 130 km, indicating that solar 
He II Lyman $ \alpha $ photons are largely degraded at higher altitudes and 
do not reach  altitudes of 130 km. The calculated flux decreases exponentially with 
increasing energy. A sharp fall in the photoelectron flux at $ \sim $70 eV is 
due to the presence of this features in the primary photoelectron energy spectrum 
resulting from photoionization process.
\citep{Jain11}.

Figure~\ref{fig:pef} (bottom panel) shows the calculated photoelectron fluxes for 
solar maximum 
condition (F10.7=200) at SZA=20\dgr. Photoelectron flux features are similar to those in 
solar minimum conditions. This figure also shows the electron flux measured by the 
Pioneer Venus Orbiter Retarded Potential Analyser \citep{Spenner97}. Good agreement both in
shape and magnitude, is observed between the calculated and measured fluxes.
\cite{Spenner97}  found that the average electron fluxes do not vary much for 
SZA between 0\dgr\ and 70\dgr.

%+++++++++++++++++++++++++++++++++++++++++++++++++++++++++++++++++++++++++++++++++++++++++++

\section{Results and discussion}
\begin{figure}
\centering
\includegraphics[width=30pc]{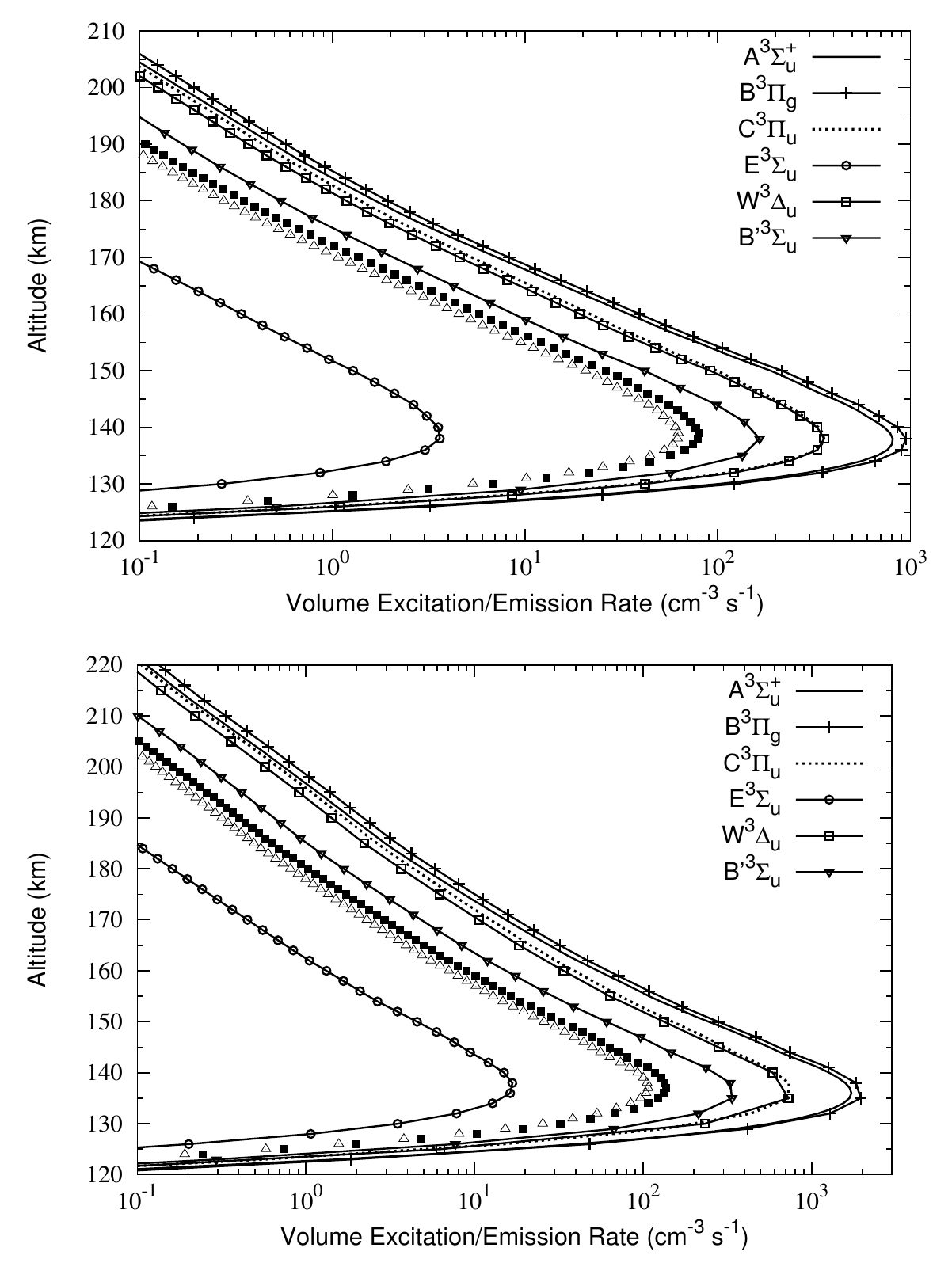}
\caption{Volume excitation rates of various 
triplet states of \nt\ by direct electron impact
excitation in solar minimum (upper panel) and maximum (bottom panel) conditions at
SZA = 60\dgr. The rate of E$^3\Sigma_u $ state is plotted after multiplying by 2. 
Solid square and open triangle represent the volume emission rates of the VK (0, 6)
and VK (0, 5) bands, respectively.}
\label{fig:n2ver}
\end{figure}

The volume excitation rate for \nt\ emissions is calculated
as 

\begin{equation}\label{eq:e}
V_i(Z, E) = n(Z) \int _{E_{th}}^{E} \phi(Z, E) \sigma_i(E) dE,
\end{equation}
where $n(Z)$ is the density of \nt\ at altitude $ Z $ and 
$ \sigma_i(E)$ is the electron impact cross section for the $i$th state, 
for which the threshold is $E_{th}$. Figure~\ref{fig:n2ver} shows 
the excitation rates of \nt\ triplet states ($ A,\, B,\, C, \, W, \,B',$ and $ E $) 
by photoelectron impact excitation. During both solar minimum and maximum 
conditions, excitation rate for all the states peaks at $ \sim $135-140 km. However,
at the peak, the excitation rates during solar maximum are a factor of 2 
higher than that during solar minimum, while at higher altitudes
($ \sim $200 km and above) the increase is by a factor of 5 or more.

All transitions between the triplet states of \nt\ and the ground state are spin
forbidden, therefore excitation of these states is primarily due to the electron
impact. The higher lying states $C$, $W$, and  $ B'$
populate the $B$ level, which in turn radiate to the $A$ level. Inter-system 
cascading $ B^3\Pi_g \rightleftharpoons A^3\Sigma^+_u $ and $ B^3\Pi_g 
\rightleftharpoons W^3\Delta_u $ are also important in populating the $ B $ 
level \citep{Cartwright71, Cartwright78}. All excitations to the higher triplet
states eventually cascade into the $ A^3\Sigma^+_u$ state
\citep{Cartwright71,Cartwright78}. The effect of reverse first positive transition
is important in populating the lower vibrational levels of the $B$ state, which in
turn populate the lower vibrational levels of the $A$ state \citep{Sharp71,Cartwright71,
Cartwright78}. The width and shape of VK bands are quite sensitive to the 
rotational temperature, making it a useful tool to monitor the neutral temperature of 
the upper atmosphere \citep{Broadfoot97}.
To calculate the contribution of cascading from higher triplet states
and interstate cascading between different states, we solve the
equations for statistical equilibrium based on the formulation
of \cite{Cartwright78}.  At a specified altitude, for a 
vibrational level $ \nu $  of a state $ \alpha $, the population is 
determined using the statistical equilibrium equation
\begin{equation}\label{eq:sta}
V^\alpha q_{0\nu}  + \sum\limits_{\beta}\sum\limits_{s}A^{\beta\alpha}_{s\nu}\, n^\beta_s
= \{ K^\alpha_{q\nu} + \sum\limits_{\gamma}\sum\limits_{r}A^{\alpha\gamma}_{\nu r} \}n^\alpha_{\nu}
\end{equation}
where
\begin{tabbing}
$ \alpha, \beta, \gamma $ \quad \= electronic states \kill
$ V^\alpha $ \> electron impact volume excitation rate \\ 
			\>	(cm$^{-3}$\ s$^{-1}$) of state $\alpha$;\\
$q_{0\nu}$   \> Franck-Condon factor for the  excitation \\
             \> from ground level to $\nu$ level of state $ \alpha $; \\
$A^{\beta\alpha}_{s\nu}$ \> transition probability (s$^{-1}$) from 
                         state  \\
               \>    $ \beta(s)$ to $ \alpha(\nu)$; \\	
$K^\alpha_{q\nu}$ \>  total electronic quenching frequency  \\
                 \> (s$^{-1}$) of level $\nu$ of state $ \alpha $ by the \\
                  \> all gases defined as: $\sum\limits_{l} K_{q(l)\nu}^\alpha \times n_{l} $; 
                     where, \\
                  \> $K_{q(l)\nu}^\alpha$ is the quenching rate coefficient  \\
                   \> of level $ \nu $ of $ \alpha $ by gas $ l $  of density $ n_l $;\\
$ A^{\alpha\gamma}_{\nu r} $ \> transition from level $ \nu $ of state $ \alpha $ 
                              to \\
                       \> vibrational  level $r$ of state $ \gamma $;\\
$ n $ \> density (cm$^{-3}$);\\
$ \alpha, \beta, \gamma $ \> electronic states;\\
$ s, r $ \> source and sink vibrational levels, \\
		\> respectively.
\end{tabbing}

While calculating the cascading from  $ C $ state, 
we have accounted for predissociation.
The $ C $ state predissociates approximately 
half the time (this is an average value for all 
vibrational levels of the $ C $ state; excluding $ \nu = $ 0, 1, which do not
predissociate at all) \citep[cf.][]{Daniell86}.
In Earth's airglow the \nt(A) levels get effectively quenched by atomic oxygen and 
the abundance of O increases with increase in altitude.
In the case of Venus the main atmospheric constituent \car\ does not quench the 
\nt(A) levels that efficiently because of much lower quenching rate coefficient for 
\car, but still there will be some collisional deactivation by other atmospheric
constituents of Venus. The quenching rates for different
vibrational levels of \nt\ triplet states by O, O$_2$, and \nt\ 
are adopted from \cite{Morrill96} and by \car\ and CO are 
taken from \cite{Dreyer74}.

\begin{figure}
\centering
\includegraphics[width=30pc]{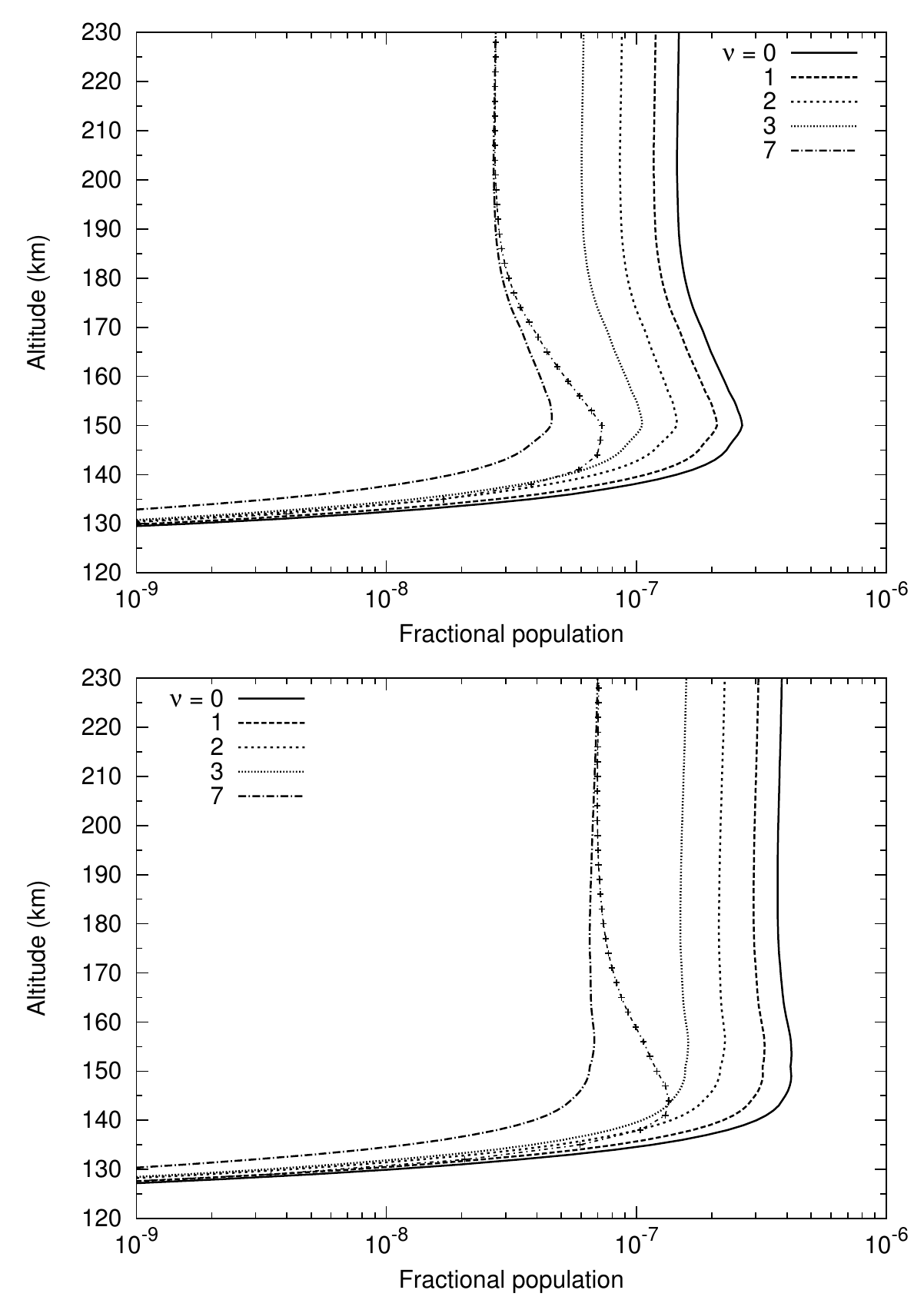}
\caption{Altitude profiles of relative population of selected vibrational
levels of the \nt(A) state with respect to the \nt(X) for low (upper panel) and high 
(bottom panel) solar
activities, at SZA=60\dgr. Dashed line with symbols shows the relative population of 
vibrational level 7 without quenching.}
\label{fig:fpop-A}
\end{figure}

\begin{figure}
\centering
\includegraphics[width=30pc]{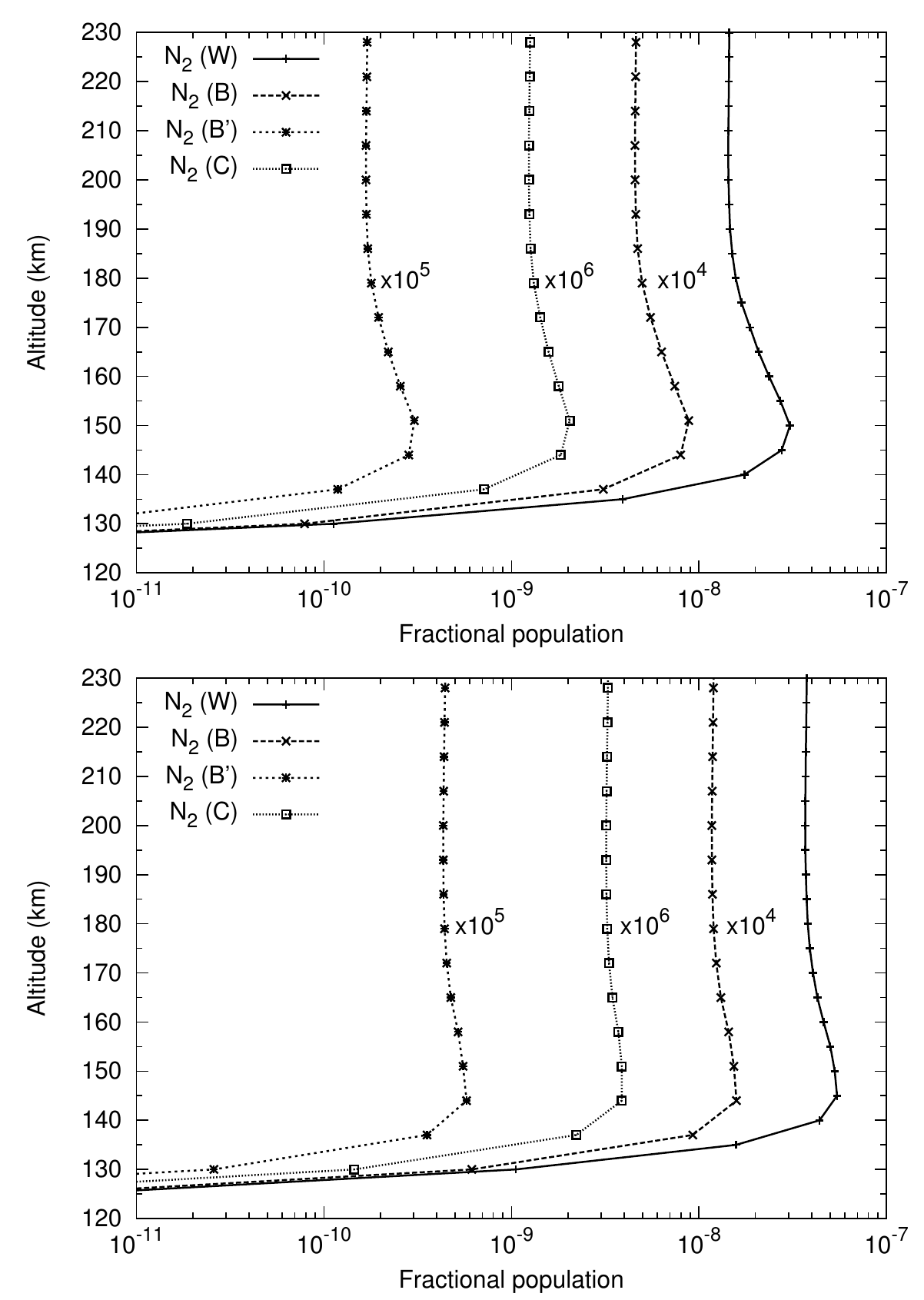}
\caption{Altitude profiles of relative population of 0 vibrational level of
$ B$, $B'$, $C$, and $W$ states of \nt\ with respect to the \nt(X) 
for low (upper panel) and high (bottom panel) solar activities, at SZA=60\dgr. Populations of
$B, B'$, and $C$ have been plotted after multiplying by a factor of 
$10^4, 10^5$, and $10^6$, respectively.}
\label{fig:fpop-B}
\end{figure}

Figure~\ref{fig:vibpop} shows the population of different 
vibrational levels of triplet states of \nt\ at 150 km relative 
to the ground state. The relative vibrational population of the \nt(A) state
at 130 km is also shown in Figure~\ref{fig:vibpop}.
Our calculated relative vibrational populations agree well 
with the  calculations of  \cite{Morrill96} and \cite{Cartwright78}.
To show the effect of quenching, the relative
vibrational populations of the \nt(A) state calculated without
quenching at 130 and 150 km are also shown in 
Figure~\ref{fig:vibpop}.
It is clear from Figure~\ref{fig:vibpop}
that quenching does affect the vibrational population of
\nt(A) state at lower altitudes,  mainly for 
vibrational levels $\leq$ 12; however, as the altitude increases the 
effect of quenching decreases.
Figure~\ref{fig:fpop-A} shows the 
steady state fractional population altitude profile of a few vibrational 
levels of the A$^3\Sigma_u^+$ excited state of \nt, while Figure~\ref{fig:fpop-B} 
shows them for $ \nu' = 0 $ level of the $ B$, $C$, $W$, and $ B' $ states
for both solar minimum and maximum conditions.
At low solar activity, the vibrational populations show a peak at $ \sim $150
km, while during high solar activity, a broad peak $ \sim $140--160 km is 
seen. The quenching reduces the population by a factor of 2 to 3 around
the peak, and its effect is felt up to an altitude of $ \sim $200 km.

After calculating the steady state density of 
different vibrational levels of excited triplet states of 
\nt, the volume emission rate $ V_{\nu'\nu''}^{\alpha\beta} $ 
of a vibration band $\nu' \rightarrow \nu''$ can be obtained using
\begin{equation}\label{eq:g}
V_{\nu'\nu''}^{\alpha\beta} = n_{\nu'}^\alpha \times 
                            A_{\nu'\nu''}^{\alpha\beta} \quad (cm^{-3}\ s^{-1})
\end{equation}
where  $n_{\nu'}^\alpha$ is the density of vibrational 
level $\nu'$ of state $ \alpha $, and $A_{\nu'\nu''}^{\alpha\beta}$
is the transition probability (s$^{-1}$) for the transition 
from the $\nu'$ level of state $ \alpha $ to the $\nu''$ level of state $ \beta $.

Volume emission rates are height-integrated to calculate the
overhead intensities. Table~\ref{tab:tableB} shows the total
overhead intensity for triplet transitions of \nt\ during low and high solar 
activity at SZA=60\dgr\ calculated by adding all the band emissions.
Table~\ref{tab:tableA} shows the nadir intensity (SZA=60\dgr) during solar minimum and
maximum conditions for the prominent VK bands of \nt; for other triplet
emissions (viz., First Positive, Second Positive, and Wu-Benesch)
overhead intensities are given in Table~\ref{tab:n2-oi}.
%In general, the intensities during high solar activity are a factor of 2 
%higher; however, in the case of VK bands the increase is 
%$\sim$60--70 \%.
The increase in the VK band intensity from solar minimum to maximum is
$\sim$60--70\%. The strongest VK band emissions (0, 5), (0, 6), 
(0, 7), (0, 8), (1, 9), and (1, 10) have intensities of
94 (160), 120 (204), 114 (194), 94 (155), and 88 (145) R, respectively, during
solar minimum (maximum) condition. On Mars the SPICAM aboard Mars Express has 
recently detected VK (0, 5), (0, 6), and (0, 7) bands \citep{Leblanc06,Leblanc07}.

The calculated band emission rate can be integrated along the line of 
sight at a projected distance from the centre of Venus to obtain limb 
profiles. Figure~\ref{fig:limb06} shows the limb profiles of VK (0, 6)
band at SZA of 0\dgr\ and 60\dgr\ for solar minimum and maximum conditions.
In the case of solar minimum, the limb profile peaks at 133 (136) km 
for SZA=0\dgr\ (60\dgr), with a value of $ \sim $10 ($\sim $5) kR. During
high solar activity, peak value of $ \sim $20 ($ \sim $10) kR in 
the profiles occurs at 131 (134 km) for SZA=0\dgr\ (60\dgr). Thus, the intensity
increases by a factor of 2 at the peak for a low to high solar activity change.
The effect of the SZA is observed at lower altitudes ($<$160 km) only.
Above 150 km the shape of the limb profile in minimum and maximum condition are different,
which is due to the difference in the Venus model atmosphere in the two solar conditions.
The limb intensities of the VK (0, 4), (0, 5), and (0, 7) bands are also calculated in the
model. The shape of the limb profiles of the VK (0, 4), (0, 5), and (0, 7) bands are
similar to that of the VK (0, 6) band shown in Figure~\ref{fig:limb06}, but their
limb intensity at the peak is around 44, 78, and 95\%, respectively, of the limb
intensity of the VK (0, 6) band.

On Mars, the SPICAM has observed limb profiles of (0, 5) and (0, 6) bands of VK
emissions, since they are the strong emissions below 3000 \AA\
(see Table~\ref{tab:tableA}). The peak
intensity of VK (0, 6) band on Mars is $\sim $0.5 kR \citep{Leblanc07,Jain11}.
This value is more than a factor of 10 lower than that calculated on Venus
during low solar activity. Only a factor of 5 difference in intensity is
expected due to the difference in heliocentric distances of Mars and Venus. A much 
larger difference is due to the higher \nt\ density around the peak on Venus (138 km, 
$ 8 \times 10^9 $ cm$^{-3}$) compared to that on Mars (126 km, $ 1.5 \times 10^9 $ cm$^{-3}$);
the \nt/\car\ ratio being 0.12 on Venus and 0.03 on Mars at the intensity peak.
However, at higher altitudes the density of \nt\ on Mars is larger compared to that
on Venus; {\it e.g.}, at 200 km, the \nt\ density at Mars and Venus is $7.5 \times 10^6$
and $ 2.4 \times 10^6 $ cm$^{-3}$, respectively.

\cite{Jain11} have made a detailed study about the effect of electron impact cross 
sections of \nt\ on the calculated VK band intensities on Mars. Their study suggests that
the use of different electron impact cross sections can change the calculated 
VK band intensities up to 50\%. The effect of using e-\nt\ cross sections
of \cite{Johnson05} on the calculated overhead intensities of total VK, 
First Positive, Second Positive, and Wu-Benesch bands on Venus is shown in
Table~\ref{tab:tableB}; intensities are smaller due to the lower cross sections
of \cite{Johnson05}.

In their study \cite{Jain11} found that use of SOLAR2000 model of \cite{Tobiska00},
instead of EUVAC model of \cite{Richards94} as the solar EUV input flux in 
the model can increase the calculated intensities of \nt\ VK bands by $ \sim $15\%
on Mars. A similar effect is found on Venus if the SOLAR2000 model of 
\cite{Tobiska00} is used.

We have also calculated intensities for triplet transitions by taking the model
atmosphere as used in \cite{Fox81} for low solar activity at SZA = 45\dgr.
Our calculated intensities for the VK bands are about 30\% higher 
\citep[similar results have been seen in the case of Mars also; cf.][]{Jain11} 
than those of \cite{Fox81}, except for the VK (0, 2) band, which is 
a factor of $ \sim $8 lower than the value of \cite{Fox81}. Similar difference 
has been observed in the overhead intensity of VK (0, 2) band on Mars
\citep{Jain11}. The factor which controls the intensity of VK (0, 2) band
is the transition probability for that band whose value is  $3.54 \times 10^{-3}$ in 
our model, taken from \cite{Gilmore92}. \cite{Piper93} has also reported
similar value for the VK (0, 2) band transition probability ($\sim4.0\times10^{-3}$).

Differences between our and \cite{Fox81} calculated overhead intensities 
of First Positive bands ($ B \rightarrow A $) are between 10 to 50\%; but for a
few bands, {\it e.g.,} (2, 0), (2, 1), (3, 1), (4, 2), and (5, 3), our calculated 
intensities are a factor of 2 to 4 higher. For Second Positive bands ($ C \rightarrow B $) 
our values for transitions from 0 level of $C$ state, are $\sim$50\% higher than 
the values of \cite{Fox81}  but for transitions from level 1, our values are 
$\sim$30\% smaller than \citeauthor{Fox81} values. For Wu-Benesch ($W - B$) band our 
calculated values are smaller than those of \cite{Fox81} by factor of 2 to 4.
These differences are mainly due to updated transition probabilities used in 
our model which we have taken from \cite{Gilmore92}. \cite{Piper93} has 
reported the transition probabilities of the VK bands, which are consistent with 
those of \cite{Gilmore92}.

\section{Summary}
\begin{figure}
\centering
\includegraphics[width=30pc]{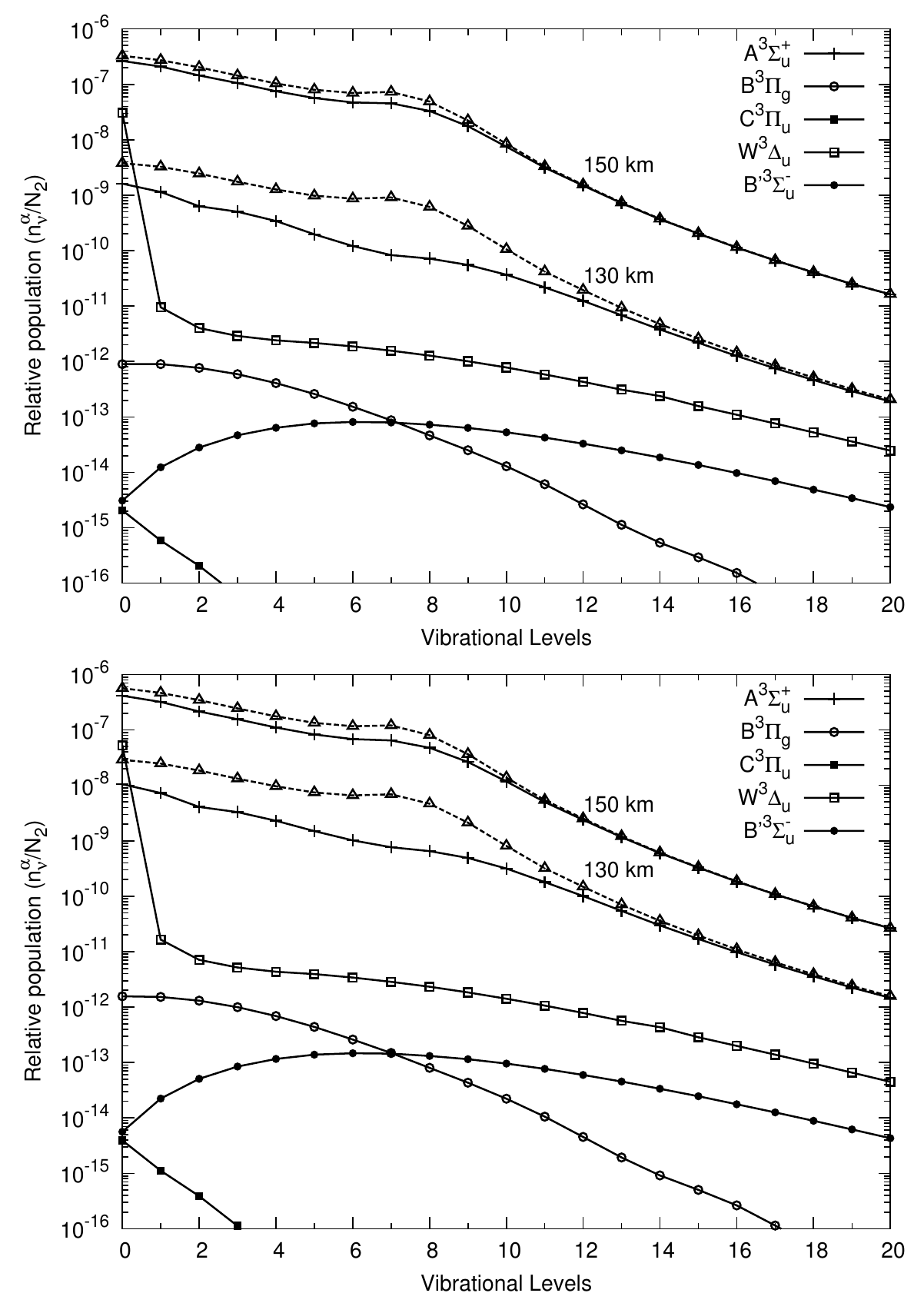}
\caption{Relative population of vibrational levels of 
different triplet state of \nt\ with respect 
to \nt (X) at 150 km for low (upper panel) and high (bottom panel) solar activities, at SZA=60\dgr.
Dashed line with triangle shows the 
relative vibrational population of  A$ ^3\Sigma_u^+$ without quenching at 
130 and 150 km.}
\label{fig:vibpop}
\end{figure}

\begin{figure}
\centering
\includegraphics[width=30pc]{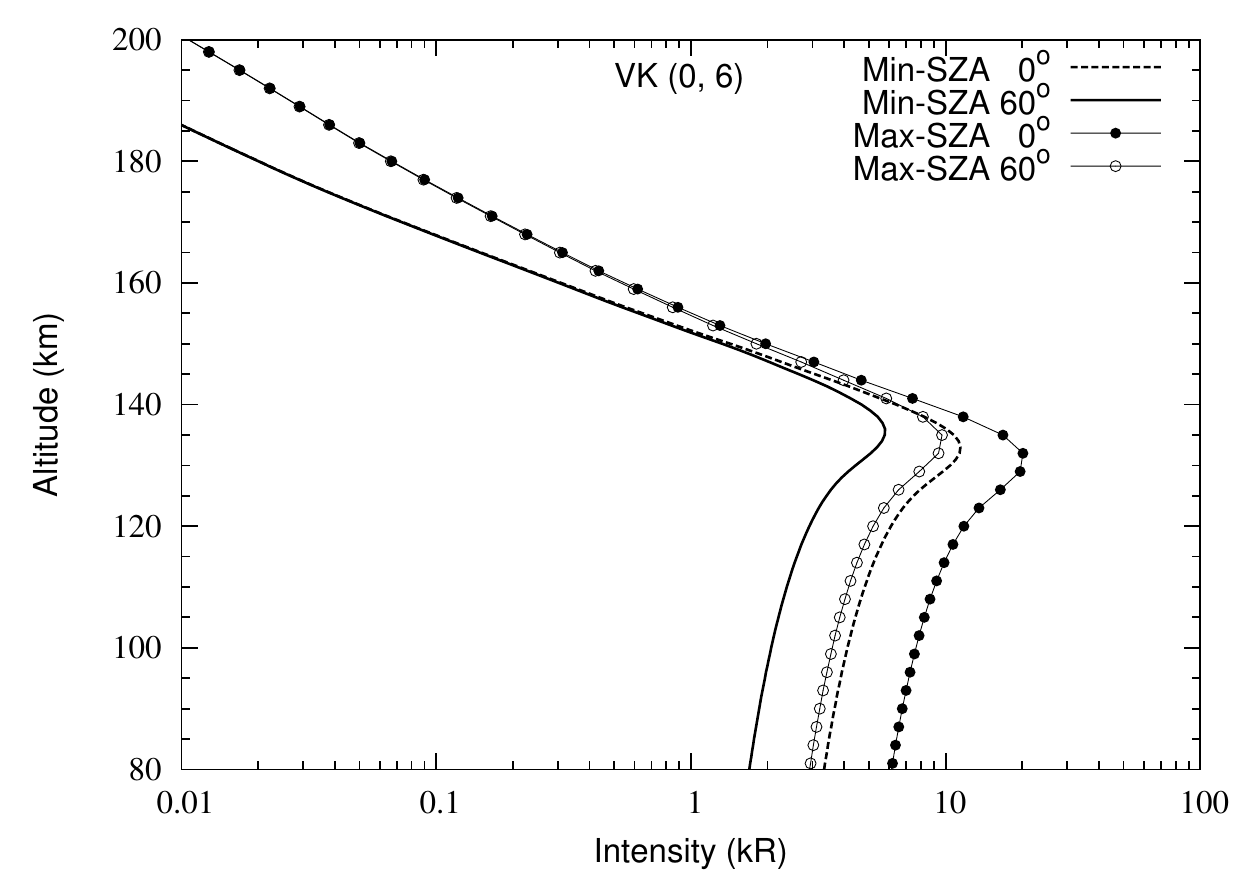}
\caption{Calculated limb intensity of VK (0, 6) band for low (min) and 
high (max) solar activity conditions, at SZA of  0\dgr\ and 60\dgr.}
\label{fig:limb06}
\end{figure}

A model for \nt\ triplet bands dayglow emissions on Venus has been 
developed for solar minimum and maximum conditions. The Analytical Yield 
Spectra technique is used to calculate the steady state 
photoelectron flux, which is in good agreement with the Pioneer Venus Orbiter-observed 
electron flux. Profiles of volume excitation rates of \nt\ VK 
bands and other triplet states are calculated using photoelectron flux.
Population of any given level of triplet states has been calculated considering 
direct electron impact excitation and quenching as well
as cascading from higher triplet states in statistical 
equilibrium conditions. The relative vibrational populations of
different triplet states have been calculated and shown 
in Figure~\ref{fig:vibpop}. Overhead intensities of Vegard-Kaplan, 
First Positive, Second Positive, Wu-Benesch, $ B' \rightarrow B $, 
$ E \rightarrow C$, $ E \rightarrow B $, $ E \rightarrow A $, and 
Reverse First Positive bands of \nt\ have been calculated and presented in 
Table~\ref{tab:tableB}. The height-integrated overhead  
intensities of prominent transitions in VK, First Positive, Second Positive,
and $ W \rightarrow B $ bands have been calculated and 
are given in Tables~\ref{tab:tableA} and \ref{tab:n2-oi}.
We have presented the calculated limb profiles of the VK (0, 6) band
for solar minimum and maximum conditions in Figure~\ref{fig:limb06}.
Maximum change in VK bands intensity are $ \sim $50\% due to changes 
in the \nt\ triplet states electron impact excitation cross sections.
An increase of $ \sim $15\% in the calculated intensities is observed when 
the SOLAR2000 solar EUV flux model of \cite{Tobiska00} is used 
instead of the EUVAC model of \cite{Richards94}.
The calculated intensities of VK bands on Venus are an order of magnitude larger
than those on Mars. Hence, the intensities are quite large and can be 
detected by the SPICAV experiment on board the Venus Express mission.
However, very bright sunlit limb due to solar scattering background makes it
difficult to observe \nt\ VK bands in Venus dayglow by SPICAV.
The relative population of vibrational levels is almost constant above 180 km
on Venus, while on Mars they attain a constant value above 250 km.

%The results presented in this paper will motivate a search of \nt\ 
%triplet emissions in the SPICAV data and would be helpful in their 
%interpretation.

%% Using an acknowledgements command is not in the Elsevier template,
%% but it can be used.
%\ack
%This work has made use of NASA's Astrophysics Data System.  It 
%also benefitted tremendously from 
%\label{lastpage}

% Bibliographic references with the natbib package:
% Parenthetical: \citep{Bai92} produces (Bailyn 1992).
% Textual: \citet{Bai95} produces Bailyn et al. (1995).
% An affix and part of a reference:
%   \citep[e.g.][Ch. 2]{Bar76}
%   produces (e.g. Barnes et al. 1976, Ch. 2).-

%\bibliography{../../references.bib}

%% Use the plainnat style for ``Icarus'' mode to display DOI numbers
%% among other things.  However, revert to the Elsevier elsart-harv
%% mode for ``Elsevier'' mode.
%\bibliographystyle{plainnat}
%\bibliographystyle{elsart-harv}
%\bibliographystyle{elsarticle-harv}
%\bibliographystyle{elsarticle-harv-doi}

\end{multicols}

\label{lastpage}

%% --Tables-- 
\clearpage	% Make sure things don't run together.
\renewcommand{\thefootnote}{\fnsymbol{footnote}}

\begin{center}
\begin{table}
\caption{Height-integrated overhead intensities of triplet transitions of \nt.}
%\small
\begin{tabular*}{\textwidth}{@{\extracolsep{\fill}}lll}
\hline
\multirow{2}{3cm}{Band} & \multicolumn{2}{c}{Intensity (kR)} \\
\cline{2-3}
		& Min\footnotemark[1]	&
		Max.\footnotemark[2]\\
\hline
Vegard-Kaplan ($A \rightarrow X$)		& 1.9 (1.5) 	 & 3.2	\\
First Positive ($B \rightarrow A$)		& 3   (2.0) 	 & 6	\\
Second Positive ($C \rightarrow B$)	& 0.4 (0.1) 	 & 0.8	\\
Wu-Benesch ($W \rightarrow B$)						& 0.5 (0.4) 	 & 1.1	\\
$B' \rightarrow B$ 					& 0.2 (0.08)	 & 0.5	\\
$E \rightarrow A$						& 3E-3 (3E-3)\footnotemark[3]	 & 7E-3\\
$E \rightarrow B$						& 5E-4 (5E-4)    & 1E-3	\\
$E \rightarrow C$						& 2E-3 (2E-3)	 & 4E-3\\
R1P\footnotemark[4] ($A \rightarrow B$)	& 0.5	(0.4)	 & 0.9	\\
\hline
\end{tabular*}

\footnotemark[1]{\small Solar minimum (F10.7=80). Values in the
bracket is for e-\nt\ cross sections taken from \cite{Johnson05}.}\\
\footnotemark[2]{\small Solar maximum (F10.7=200).}\\
\footnotemark[3]{\small 3E-3 = $ 3 \times 10^{-3}$.}\\
\footnotemark[4]{\small Reverse First Positive.}
\label{tab:tableB}
\end{table}
\end{center}

{
\renewcommand{\baselinestretch}{1}
\small\normalsize
\begin{center}
\begin{table}
\caption{ \nt\ Vegard-Kaplan Band ($ A^3\Sigma^+_u \rightarrow X^1\Sigma^+_g $)
height-integrated overhead intensity.}
%\small
\begin{tabular*}{\textwidth}{@{\extracolsep{\fill}}llcc}
\hline \noalign{\smallskip}
\multirow{2}{1.2cm}{Band $\nu'-\nu''$} & Band & \multicolumn{2}{c}{Overhead Intensity (R)}\\
\cline{3-4}
            & \multicolumn{1}{c}{Origin} & \multicolumn{1}{l}{Min.\footnotemark[1]} &\multicolumn{1}{c}{Max.\footnotemark[3]}\\ \noalign{\smallskip}
\hline \noalign{\smallskip}
0-2	& 2216	& 4    & 7   \\
0-3	& 2334	& 20   & 34  \\
0-4	& 2463	& 53   & 91  \\
0-5 & 2605 	& 94   & 160 \\
0-6	& 2762	& 120  & 204 \\
0-7 & 2937	& 114  & 194 \\
0-8	& 3133	& 84   & 143 \\
0-9	& 3354	& 49   & 84  \\
1-3	& 2258	& 16   & 27  \\
1-4	& 2379	& 24   & 40  \\
1-8	& 2998	& 63   & 105 \\
1-9	& 3200	& 94   & 155 \\
1-10& 3427	& 88   & 145 \\
1-11& 3685	& 59   & 97  \\
1-12& 3980	& 30   & 49  \\
1-13& 4321 	& 12   & 19  \\
2-10& 3270	& 26   & 41  \\
2-11& 3503	& 54   & 86  \\
2-12& 3769	& 58   & 93  \\
2-13& 4074	& 41   & 66  \\
3-12& 3583 	& 15   & 24  \\
3-13& 3857	& 37   & 60  \\
3-14& 4171	& 41   & 67  \\
4-11& 3198	& 16   & 26  \\
4-15& 4274	& 25   & 41  \\
4-16& 4650	& 26   & 42  \\
5-17& 4771	& 18   & 29  \\
5-18& 5229	& 16   & 26  \\
6-19& 5372	& 13   & 21  \\
7-0	& 1689	& 11   & 18  \\
8-0	& 1655	& 12   & 20  \\
9-0	& 1622	& 9    & 16  \\
\hline
\end{tabular*}

\footnotemark[1]{\small Solar minimum condition.} \\
\footnotemark[3]{\small Solar maximum condition.}
\label{tab:tableA}
\end{table}
\end{center}
}

{
\renewcommand{\baselinestretch}{0.6}
\small\normalsize
\begin{center}
\begin{table}
\caption{Calculated height-integrated overhead intensity of 
\nt\ triplet emissions.}
\begin{tabular*}{\textwidth}{@{\extracolsep{\fill}}ccccccccc}
\hline \noalign{\smallskip}
Band & Band Origin & \multicolumn{2}{c}{Intensity (R)} &
&Band & Band Origin & \multicolumn{2}{c}{Intensity (R)} \\ 
%Band & Band Origin & \multicolumn{2}{c}{Intensity (R)}  \\
\cline{3-4}  \cline{8-9}
($ \nu' - \nu''$) & \AA\ & Min.\footnotemark[1] & Max.\footnotemark[2]& &
($ \nu' - \nu''$) & \AA\ & Min.\footnotemark[1] & Max.\footnotemark[2] \\ \noalign{\smallskip}
%($ \nu' - \nu''$) & \AA\ & Min.\footnotemark[1] & Max.\footnotemark[2]\\
%\hline \noalign{\smallskip}
\cline{1-4}  \cline{6-9} \noalign{\smallskip}
\multicolumn{4}{c}{First Positive  $B^3\Pi_g$-- $A^3\Sigma^+_u$} &
&\multicolumn{4}{c}{Reverse first positive ($A^3\Sigma_u^+$-- $B^3\Pi_g$)} \\ [5pt]
0-0	&	10469	& 	229     &	465 && 9-0	&	42700	&	29  &	51   \\
0-1	&	12317 	& 	123   	&	250 && 10-1  &	55100	&	45  &	84   \\
0-2	&	14895	&	37  	&	75  && \\
1-0	&	8883	&	367  	&	744 &&\multicolumn{4}{c}{Herman-Kaplan ($E^3\Sigma_g^+$-- $A^3\Sigma_u^+$)} \\ \noalign{\smallskip}
1-2	&	11878	&	68  	&	138 && 0-1   &   2243	&	0.35 & 0.8   \\
1-3	&	14201	&	52  	&	105 && 0-2   &   2316	&	0.57 & 1.3   \\
2-0	&	7732	&	182 	&	372 && 0-3   &   2392	&	0.63 & 1.4   \\
2-1	&	8695	&	236 	&	480 && 0-4   &   2472	&	0.52 & 1.2   \\
2-2	&	9905	&	44  	&	89   \\
2-4	&	13572	&	33  	&	68  &&\multicolumn{4}{c}{$E^3\Sigma_g^+$-- $B^3\Pi_g$} \\ \noalign{\smallskip}
3-1	&	7606	&	252 	&	515 && 0-1   &   2877	&	0.14 & 0.3   \\
3-2	&	8516	&	66  	&	135 && 0-2   &   3181	&	0.11 & 0.3   \\
3-3	&	9648	&	79  	&	162  \\
4-1	&	6772	&	77  	&	158 &&\multicolumn{4}{c}{$E^3\Sigma_g^+$-- $C^3\Pi_u$} \\ \noalign{\smallskip}
4-2	&	7484	&	193  	&	395 && 0-0   &   14713	&	1.7  & 3.8   \\
4-4	&	9404	&	58  	&	119 && 0-1   &   20824	&	0.14 & 0.3   \\
5-2	&	6689	&	87  	&	178  \\
5-3	&	7368	&	103  	&	211 &&\multicolumn{4}{c}{$B'^3\Sigma_u^-$-- $B^3\Pi_g$} \\ \noalign{\smallskip} 
6-3	&	6608	&	71  	&	145 && 3-1   &   10816	&	5.4  & 11.2 \\
7-4	&	6530	&	47  	&	96  && 4-1   &   9375	&	6.3  & 13   \\ \noalign{\smallskip}
\multicolumn{4}{c}{Wu-Benesch ($W^3\Delta_u$-- $B^3\Pi_g$)}&	& 4-2   &   8990	&	5.8  & 12.2 \\\noalign{\smallskip}
2-0	&	33206	&	16   	&	34  && 5-2   &   9626	&	9.1  & 19   \\
3-0	&	22505	&	14  	&	29  && 6-2   &   8501	&	8    & 16.9 \\
3-1	&	36522	&	12   	&	25  && 6-3   &   9886	&	8.7  & 18.2 \\
4-1	&	24124	&	21  	&	43  && 8-4   &   8947	&	8.6  & 18   \\
5-1	&	18090	&	19  	&	40  && 9-4   &   8011	&	7.5  & 15.6 \\
5-2	&	25962	&	18  	&	38  && 9-5   &   9180	&	6.3  & 13   \\
6-2	&	19193	&	23  	&	49  && 10-4  &   7264	&	4.8  & 10.1 \\
7-2	&	15281	&	19  	&	39  && 10-5  &   8213	&	6.5  & 13.6 \\ \noalign{\smallskip}
7-3	&	20421	&	21  	&	43  && \multicolumn{4}{c}{Second Positive $C^3\Pi_u$-- $B^3\Pi_g$} \\ \noalign{\smallskip}
8-3	&	16112	&	21  	&	43  && 0-0	&	3370	&	137 	&	291 \\ 
9-3	&	13347	&	15 		&	31	&& 0-1	&	3576	&	92  	&	196 \\ 
9-4	&	17024	&	18 		&	37	&& 0-2	&	3804	&	37  	&	79  \\ 
10-4&	14014	&	15	 	&	31	&& 1-0	&	3158	&	35      &   75  \\ 
10-5&	18030	&	12.5 	&	26	&& 1-2	&	3536	&	16  	&   35  \\ 
\hline
\end{tabular*}

\footnotemark[1]{\small Solar minimum condition.}\\
\footnotemark[2]{\small Solar maximum condition.}
\label{tab:n2-oi}
\label{lasttable}
\end{table}
\end{center}
}

\clearpage

%% --Figures-- %%

\end{document}